\documentclass[prd,aps,10pt,nofootinbib,twocolumn,superscriptaddress,preprintnumbers,balancelastpage,longbibliography]{revtex4-1}

\usepackage{dcolumn}
\usepackage{xcolor}
\usepackage{youngtab}
\usepackage{ytableau}

\usepackage{booktabs}    
\usepackage{placeins}
\usepackage{soul}

\definecolor{redd}{rgb}{0.8, 0.1,0.2}
\definecolor{navy}{rgb}{0.05, 0.23,0.75}
\usepackage[colorlinks=true,citecolor=red,linkcolor=blue]{hyperref}
\usepackage{chngcntr}
\hypersetup{
     colorlinks   = true,
     citecolor    = navy,
	linkcolor = redd,
	urlcolor=navy,
	anchorcolor=blue
}
\usepackage{bbm}
\usepackage{amsfonts}
\usepackage{amsmath,amssymb}
\usepackage{mathrsfs}
\usepackage{epsfig}
\usepackage{graphicx}
\usepackage{wrapfig}                 
\usepackage{url}
\usepackage{hyperref}
\usepackage{float}
\usepackage{color}
\usepackage{multirow}
\usepackage{lipsum}
\usepackage{enumitem}
\usepackage{ctable}
\newcolumntype{L}{>{\centering\arraybackslash}m{1.5cm}}

\usepackage{enumitem}
\usepackage{chngcntr}
\newcommand{\nn}{\nonumber}

\newcommand{\be}{\begin{equation}}
\newcommand{\ee}{\end{equation}}
\newcommand{\bea}{\begin{eqnarray}}
\newcommand{\eea}{\end{eqnarray}}
\newcommand{\bc}{\begin{center}}
\newcommand{\ec}{\end{center}}

\begin{document}
		
\title{A novel strategy to prove chiral symmetry breaking in QCD-like theories}
	
\author{Luca Ciambriello}
\email{luca.ciambriello@unicatt.it}
\affiliation{Interdisciplinary Laboratories for Advanced Materials Physics (i-LAMP) and Dipartimento di Matematica e Fisica, Universit\`{a} Cattolica del Sacro Cuore, Brescia, Italy}

\author{Roberto Contino}
\email{roberto.contino@uniroma1.it}
\affiliation{Dipartimento di Fisica, Sapienza Universit\`{a} di Roma, Italy}
\affiliation{Istituto Nazionale di Fisica Nucleare (INFN), Sezione di Roma, Italy}

\author{Andrea Luzio}
\email{andrea.luzio@sns.it}
\affiliation{Scuola Normale Superiore, Pisa, Italy}
\affiliation{Istituto Nazionale di Fisica Nucleare (INFN), Sezione di Pisa, Italy}

\author{Marcello Romano}
\email{marcello.romano@ipht.fr}
\affiliation{Universit\'e Paris-Saclay, CNRS, CEA, Institut de Physique Théorique, 91191, Gif-sur-Yvette, France}

\author{Ling-Xiao Xu}
\email{lxu@ictp.it}
\affiliation{Abdus Salam International Centre for Theoretical Physics, Trieste, Italy}

\begin{abstract}
  We demonstrate that chiral symmetry breaking occurs in the confining regime of QCD-like theories with $N_c$ colors and $N_f$ flavors.
  Our proof is based on a novel strategy, called `downlifting', by which solutions of the 't Hooft anomaly matching and persistent mass conditions for a theory with $N_f-1$ flavors are constructed from those of a theory with $N_f$ flavors, while $N_c$ is fixed. By induction, chiral symmetry breaking is proven for any $N_f\geq p_{min}$ in the confining regime, where $p_{min}$ is the smallest prime factor of $N_c$. The proof can be extended to $N_f <p_{min}$ under the additional assumption on the absence of phase transitions when quark masses are sent to infinity. Our results do not rely on assumptions on the spectrum of massless bound states other than the fact that they are color-singlet hadrons.
\end{abstract}

\maketitle
	
\section{Introduction}
One of the long-standing challenges in Quantum Field Theory (QFT) and particle physics is to understand color confinement and the spontaneous breaking of global symmetries in strongly-coupled four-dimensional gauge theories like Quantum Chromodynamics (QCD).
Thanks to asymptotic freedom~\cite{Gross:1973id, Politzer:1973fx}, QCD at high energy is described in terms of weakly-coupled quarks and gluons as fundamental degrees of freedom; its low-energy degrees of freedom,
on the other hand, are color-singlet bound states of the underlying strong dynamics.
While numerical methods have been successful to characterize individual theories~\cite{Wilson:1974sk, Kogut:1974ag, Susskind:1976jm}, strongly-coupled four-dimensional QFTs are not amenable to direct analytical control, with the notable exceptions of supersymmetric theories.
For this reason, demonstrating color confinement and global symmetry breaking analytically is a challenging task and has remained an unsolved problem so far.

Arguments based on 't Hooft anomaly matching have been put forward in the literature to prove, in QCD-like theories with generic numbers of colors $N_c$ and flavors $N_f$, the absence of a low energy confining regime with unbroken chiral symmetry.~\footnote{This has to be contrasted with ${\cal N}=1$ supersymmetric QCD for $N_f=N_c+1$~\cite{Seiberg:1994bz}, see also~\cite{Csaki:1996sm,Csaki:1996zb} for other supersymmetric gauge theories. The same phenomenon is considered possible also in non-supersymmetric gauge theories, see for example~\cite{Bars:1981se, Poppitz:2019fnp}.}
Such arguments however rely on dynamical assumptions on the spectrum of massless bound states, see~\cite{Ciambriello:2022wmh} and the discussion below. In this paper, we provide a proof of chiral symmetry breaking ($\chi$SB) for generic QCD-like theories that holds true for a number of flavors $N_f\geq p_{min}$, where $p_{min}$ is the smallest prime factor of the number of colors $N_c$.  Our only assumption is that the theory confines.
We also provide an argument to extend the proof to $N_f < p_{min}$ at the cost of assuming the absence of phase transitions when the quark masses are sent to infinity.

A QCD-like theory, dubbed QCD$[N_c, N_f]$ in the following, has $N_f$ flavors of vectorlike quarks in the fundamental representation of the gauge group $SU(N_c)$. When all the quarks are massless, its global symmetry group is
\bea
\mathcal{G}[N_f]=\frac{SU(N_f)_L\times SU(N_f)_R\times U(1)_V}{\mathbb{Z}_{N_c} \times \mathbb{Z}_{N_f}}\, ,
\label{eq:G}
\eea
for $N_c\geq 3$ and $N_f\geq 2$.
See for example~\cite{Tanizaki:2018wtg} and the Supplementary Material for explanations on the discrete quotient.
It is believed that upon confinement the chiral symmetry $SU(N_f)_L\times SU(N_f)_R$ is spontaneously broken down to its vectorial subgroup $SU(N_f)_V$.
This wisdom is supported by results from lattice simulations~\cite{Engel:2014cka, Engel:2014eea}, QCD inequalities~\cite{Weingarten:1983uj}, and, most importantly, the existence of pions in nature, which is the hallmark of $\chi$SB in QCD~\cite{Nambu:1961tp, Nambu:1961fr}. 

\emph{How to analytically demonstrate $\chi$SB in the confining regime of QCD-like theories?} It is possible to address this question thanks to  the seminal work of 't Hooft~\cite{tHooft:1979rat}, where it is shown that the 't Hooft anomaly of the global symmetry group $\mathcal{G}[N_f]$ must be the same when computed in terms of quarks and in terms of hadrons of the low-energy effective description
(see~\cite{Frishman:1980dq, Coleman:1982yg} for a derivation of this result from unitarity and analyticity). If chiral symmetry is spontaneously broken, then the anomaly of quarks in the ultraviolet (UV) is matched by that of Nambu-Goldstone bosons in the infrared (IR). In the case of unbroken chiral symmetry, on the other hand, massless spin-1/2 fermions must exist in representations of $\mathcal{G}[N_f]$ whose multiplicities satisfy a set of consistency conditions, known as 't Hooft anomaly matching conditions (AMC). One can thus prove $\chi$SB by showing that no solution of the AMC exists for \emph{any} possible spectrum of massless fermions. 
However, AMC alone are not in general sufficient to prove $\chi$SB in QCD-like theories at finite $N_c$~\cite{Preskill:1981sr},~\footnote{See~\cite{Coleman:1980mx} for a proof of $\chi$SB in the large-$N_c$ limit.}  it is necessary to impose another set of equations called persistent mass conditions (PMC). They were formulated in~\cite{Preskill:1981sr} starting from a corresponding set of decoupling conditions introduced by 't Hooft in~\cite{tHooft:1979rat}. PMC arise from giving positive-definite masses to the quarks, and can be rigorously derived from the Vafa-Witten theorem~\cite{Vafa:1983tf}, see~\cite{Ciambriello:2022wmh}.
One can thus aim at finding a general proof of $\chi$SB by combining AMC and PMC. Previous attempts in the literature, see~\cite{tHooft:1979rat, Frishman:1980dq, Farrar:1980sn, Cohen:1981iz, Schwimmer:1981yy, Kaul:1981fd, Takeshita:1981sx}, are however based on arguments which rely on additional assumptions on the putative spectrum of massless fermions~\cite{Ciambriello:2022wmh}.

In this Letter, we demonstrate $\chi$SB in confining QCD$[N_c, N_f]$ by adopting a novel strategy of using AMC and PMC. 
The argument combines induction and contradiction (see Fig. \ref{fig:strategy} for a cartoon):
\begin{figure}[t]
\centering
\includegraphics[scale=0.85]{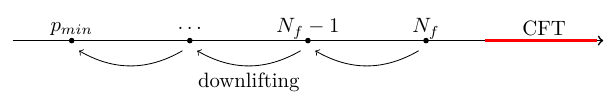}
\caption{A cartoon that summarizes our proof. Start from a value of $N_f$ where the theory confines (i.e. a value below the lower edge of the conformal window, in red), and assume that there is no chiral symmetry breaking. This implies that the spectrum of massless composite fermions solves the AMC and PMC for $N_f$ flavors. Then construct new solutions of AMC and PMC for $N_f'<N_f$, by applying the downlifting algorithm. Finally, land to $N'_f=p_{min}$, where AMC has no solution, leading to contradiction.}
\label{fig:strategy}
\end{figure}
\begin{enumerate}
\item If QCD$[N_c, N_f]$ confines without breaking the chiral symmetry, the multiplicities of massless fermions in the spectrum are a solution of the AMC and PMC for $N_f$ flavors. 
 \item Given an integer solution of AMC and PMC for $N_f$ flavors, it is possible to construct an integer solution (the \emph{`downlifted'} solution) of the AMC and PMC for $N_f-1$ flavors.
By iteration, one obtains integer solutions of AMC and PMC for any $2\leq N_f'\le N_f$.
 \item We prove that AMC for $N_f'=mp$ do not have integral solutions, where $p$ is any prime factor of $N_c$ and $m$ is any positive integer. In particular, the AMC for $N_f'=p_{min}$ do not have integral solutions, contradicting the assumption of step 1.
 \end{enumerate}
 This implies that for $N_f\ge p_{min}$ there exists no putative spectrum of QCD$[N_c, N_f]$ that solves AMC and PMC, hence the theory cannot confine without breaking the chiral symmetry.
For $N_c=3$, like in real QCD, our argument is sufficient to prove $\chi$SB for $N_f\geq 3$ if the theory confines. We will comment on the $N_f=2$ case in the end.

\section{Low energy confining description}

Before presenting our proof, we clarify how we use the notion of confinement in this paper.

In pure $SU(N_c)$ Yang-Mills theory, the confining phase can be characterized by a linear potential between two test charges (see, e.g.,~\cite{Greensite:2011zz}) and unbroken center symmetry (a one-form symmetry, in modern terminology~\cite{Gaiotto:2014kfa}).
This should be contrasted with theories with massless quarks in the fundamental representation where the confining string  is fully screened even at very low energies. In that case, due to the absence of the center symmetry, one cannot define a confining phase in the Landau sense.

Relatedly, it is well-known that for theories lacking center symmetry, the confining regime and the Higgs regime can be smoothly connected without encountering a phase transition. This phenomenon, known as Higgs-confinement continuity (see, e.g.,~\cite{Dumitrescu:2023hbe} for discussions in modern terminology), implies that these regimes are not distinct phases but rather belong to the same phase where the string is screened.

In this Letter, by `confining theory' we mean one that admits a low-energy description in terms of color-singlet particles (hadrons) interpolated by gauge-invariant local operators.
Such particles are naturally organized into proper (i.e. non-projective) irreducible representations (irreps) of $\mathcal{G}[N_f]$.~\footnote{Notice that such a description equally applies to the Higgs regime, although the spectrum of particles may be different.}
A similar description of confinement appears in various supersymmetric gauge theories~\cite{Seiberg:1994bz, Csaki:1996sm,Csaki:1996zb}.

\section{Structure of AMC and PMC}
In QCD$[N_c, N_f]$ there are in general four (perturbative) 't Hooft anomalies, $[SU(N_f)_{L,R}]^2 U(1)_V$ and $[SU(N_f)_{L,R}]^3$, which must be matched.~\footnote{If parity is spontaneously broken by the vacuum, then the $[U(1)_V]^3$ anomaly must also be matched. We will not need to use this additional condition in our proof.}
The corresponding AMC equations for unbroken chiral symmetry have the form $A_{\text{UV}} =A_{\text{IR}}$ and will be denoted by AMC$[N_f]$, to indicate that they hold in the theory with $N_f$ flavors.
We adopt the convenient normalization (see Appendix A of~\cite{Ciambriello:2022wmh}) where $A_{UV}=N_c$, so that the UV anomaly arising from the quarks is constant and independent of $N_f$. The IR anomaly $A_{\text{IR}}$ depends instead on the spectrum of massless composite fermions.

Due to strong interactions, the dynamical formation of bound states is not under analytic control. Hence, we must consider all the possible irreps to derive a general proof. Let us denote by $\mathcal{R}[N_f]$ the space of proper irreps of $\mathcal{G}[N_f]$. Each irrep $r$ contributes to $A_{\text{IR}}$ with its individual anomaly coefficient $A(r)$ multiplied by an index $\ell(r)$:
\be
A_{\text{IR}}=\sum_{r\in \mathcal{R}[N_f]} A(r) \ \ell(r)\ .
\ee
Here $\ell(r)$ equals the number of massless fermions transforming as $r$ with helicity $+1/2$ minus the number of those with helicity $-1/2$.
Therefore, for a physical spectrum, the indices $\ell(r)$ must be integer numbers.

One can deform the massless QCD-like theory by turning on positive-definite quark masses, and in this way one obtains the PMC~\cite{Preskill:1981sr, tHooft:1979rat}. When one flavor of quarks becomes massive, the group $\mathcal{G}[N_f]$ gets explicitly broken to 
\bea
\frac{SU(N_f-1)_L\times SU(N_f-1)_R\times U(1)_{\hat{V}} \times U(1)_{H_1} }{\mathbb{Z}_{N_c} \times \mathbb{Z}_{N_f-1}} , \ \
\label{eq:G1}
\eea
which we denote as $\mathcal{G}[N_f, 1]$. According to our definition, massless quarks are charged under $U(1)_{\hat{V}}$ and the massive quark is charged under $U(1)_{H_1}$, so that the $U(1)_V$ charge in $\mathcal{G}[N_f]$ equals the sum of charges under $U(1)_{\hat{V}}$ and $U(1)_{H_1}$ in $\mathcal{G}[N_f,1]$.
Bound states of the theory with one massive flavor are classified in irreps of $\mathcal{G}[N_f, 1]$, and the latter are obtained by decomposing each $r\in \mathcal{R}[N_f]$. 
We call $\hat{\mathcal{R}}[N_f,1]$ the space of proper irreps of $\mathcal{G}[N_f, 1]$ with nonvanishing $U(1)_{H_1}$ charges, and $\mathcal{R}_0[N_f,1]$ that with vanishing $U(1)_{H_1}$ charges. Two comments are in order. 

First, the Vafa-Witten theorem~\cite{Vafa:1983tf} implies that bound states in irreps of $\hat{\mathcal{R}}[N_f,1]$ must be massive~\cite{Ciambriello:2022wmh}, hence their indices vanish. This leads to a set of PMC equations,
\be
\sum_{r\in \mathcal{R}[N_f]} \ell(r) k(r\to r^\prime)=0  \qquad \forall \, r^\prime\in \hat{\mathcal{R}}[N_f,1]\,,
\label{eq:PMC_nf_1}
\ee
which will be denoted by PMC$[N_f,1]$. The integer $k(r\to r')$ denotes how many times the irrep $r'$ appears in the decomposition of $r$. 

Second, irreps in $\mathcal{R}_0[N_f,1]$ can be put into a one-to-one correspondence with irreps in $\mathcal{R}[N_f-1]$, which is the irrep space of QCD$[N_c,N_f-1]$, i.e. the QCD-like theory with $N_c$ color and $N_f-1$ massless flavors~\cite{Ciambriello:2022wmh}. This is easily seen from the global symmetry $\mathcal{G}[N_f, 1]$ in Eq.~(\ref{eq:G1}): since $U(1)_{H_1}$ acts trivially on the irreps in $\mathcal{R}_0[N_f,1]$, it can be neglected. As a result, $\mathcal{G}[N_f, 1]$ acts in the same way as $\mathcal{G}[N_f-1]$, and $\mathcal{R}_0[N_f,1]$ is identical to $\mathcal{R}[N_f-1]$, i.e.
\be
\mathcal{R}_0 [N_f, 1] \sim \mathcal{R}[N_f-1]\; . 
\label{eq:identification}
\ee 
By turning on more quark masses (each with a different value), one can decompose the irreps in $\mathcal{R}_0[N_f,1]$ further and obtain more PMC equations. We denote by PMC$[N_f, i]$ those with $i$ massive flavors, where $2\leq i \leq N_f-2$. Furthermore, we call PMC$[N_f]$ the collection of all PMC$[N_f, i]$.
Due to the identification of Eq.~(\ref{eq:identification}), each equation in PMC$[N_f, i]$ can be identified with an equation in PMC$[N_f-1, i-1]$, i.e. 
\be
\text{PMC}[N_f, i] \sim \text{PMC}[N_f-1, i-1]
\label{eq:identification_PMC}
\ee
for $2\leq i \leq N_f-2$. Notice, however, that PMC$[N_f, 1]$ are different from PMC$[N_f-1, 1]$ for a generic spectrum of massless composite fermions, see~\cite{Ciambriello:2022wmh,Ciambriello:2024msu} for concrete examples. 

Our goal is to prove that there exists no set of integer indices $\{\ell (r)\}$ that solves both AMC$[N_f]$ and PMC$[N_f]$. This implies that $\chi$SB must occur in QCD$[N_c,N_f]$ if the theory confines.
In the following, we directly discuss our proof and refer the reader to~\cite{Ciambriello:2022wmh, Ciambriello:2024msu} for more details on the allowed irreps, the structure of PMC$[N_f]$ and a thorough discussion on the results of~\cite{tHooft:1979rat, Frishman:1980dq, Farrar:1980sn, Cohen:1981iz, Schwimmer:1981yy, Kaul:1981fd, Takeshita:1981sx}.

\section{Downlifting}
The following theorem holds true:
\\[0.25cm] \textit{Let $\{ \ell (r) \}$ be a solution of $\text{AMC}[N_f]\cup\text{PMC}[N_f]$; then $\{ \tilde\ell(r^\prime)\}$ is a solution of $\text{AMC}[N_f-1]\cup\text{PMC}[N_f-1]$ for}
\be
\tilde \ell(r^\prime) \equiv \sum_{r\in \mathcal{R}[N_f]} \ell(r) \; k\left(r \to r^\prime\right) \quad \forall \, r^\prime \in \mathcal{R}[N_f-1]\,.
\label{eq:downlifting}
\ee 
We will refer to $\{\tilde\ell(r^\prime)\}$ as the downlifted solution (see also~\cite{Ciambriello:2022wmh}).
The proof of the theorem goes as follows.

First, let us consider irreps $r^\prime\in \mathcal{R}_0[N_f,1]$: their indices in the spectrum are calculable from the decomposition of $r\in \mathcal{R}[N_f]$, i.e. 
\be
\label{eq:lrprime}
\ell(r^\prime) \equiv \sum_{r\in \mathcal{R}[N_f]} \ell(r) \; k\left(r \to r^\prime\right) \quad \forall \, r^\prime\in \mathcal{R}_0[N_f,1]\ .
\ee
Since irreps of $\mathcal{R}_0[N_f,1]$ have vanishing $U(1)_{H_1}$ charge, the $\ell(r^\prime)$ are not subject to PMC$[N_f,1]$. On the other hand, $\{ \ell (r) \}$ satisfy PMC$[N_f,i]$ with $2\leq i\leq N_f-2$, which are obtained by further decomposing $r'$. The identifications of Eqs.~(\ref{eq:identification}) and~(\ref{eq:identification_PMC}) therefore imply that $\{ \tilde \ell (r') \}$, with $\tilde \ell(r') \equiv \ell(r')$ as given by Eq.~(\ref{eq:downlifting}), automatically solve PMC$[N_f-1,i-1]$ for $2\leq i\leq N_f-2$; collectively all of these equations are just PMC$[N_f-1]$.

Second, we show that the ansatz of Eq.~(\ref{eq:downlifting}) also solves AMC$[N_f-1]$. One can evaluate the anomaly coefficients $A(r)$ of either $[SU(N_f)_{L,R}]^2 U(1)_V$ or $[SU(N_f)_{L,R}]^3$ on the $SU(N_f-1)_{L,R}$ Lie subalgebra. Following the rule of decomposition, we obtain
\be
\label{eq:decA}
A(r)=\sum_{\text{All}\ r^\prime} \ k(r \to r^\prime) \  A(r^\prime)\ ,
\ee
where the sum runs over all $r^\prime$ after decomposition, i.e. $r^\prime \in \mathcal{R}_0[N_f,1] \cup \hat{\mathcal{R}}[N_f,1]$,
and $A(r^\prime)$ is the anomaly coefficient of either $[SU(N_f-1)_{L,R}]^2 U(1)_V$ or $[SU(N_f-1)_{L,R}]^3$ for any $r^\prime$. Plugging Eq.~(\ref{eq:decA}) in AMC$[N_f]$ and switching the order of sums, we have
\bea
A_{\text{UV}}
&=&\sum_{r \in \mathcal{R}[N_f]} \ell(r) \left(\sum_{\text{All}\ r^\prime}\;  \ k(r \to r^\prime) \ A(r^\prime) \right)  \nn\\
&=&\sum_{\text{All}\ r^\prime} \left(\sum_{r \in \mathcal{R}[N_f]}\; \ell(r) \ k(r \to r^\prime)\right) \ A(r^\prime) \ . \nn\\
\eea
Furthermore, PMC$[N_f,1]$ (cf. Eq.~(\ref{eq:PMC_nf_1})) imply that the sum in the parenthesis in the second line vanishes unless $r^\prime \in \mathcal{R}_0[N_f,1]$; therefore
\be
A_{\text{UV}}
=\sum_{r^\prime\in \mathcal{R}_0[N_f,1]} \ell(r^\prime) \ A(r^\prime)\, .
\ee
By the identification in Eq.~(\ref{eq:identification}), these equations have the same form as AMC$[N_f-1]$, hence $\{ \tilde\ell(r^\prime)\}$ defined by Eq.~(\ref{eq:downlifting}) is a solution of AMC$[N_f-1]$. This completes our proof.

We end this section with a few comments. 
Downlifting crucially relies on the PMC with more than one massive flavor and on the identifications of Eqs.~(\ref{eq:identification}) and~(\ref{eq:identification_PMC}). To the best of our understanding, it is not possible to downlift a generic spectrum without these PMC (see the Supplementary Material on downlifting only baryons). 
Morally speaking, our use of PMC with more than one massive flavor is analogous to Seiberg's approach of adding holomorphic quark mass terms to decouple flavors in supersymmetric QCD~\cite{Seiberg:1994bz, Seiberg:1994pq}: PMC are consistency conditions and allow us to reach the theory with $N_f-1$ flavors from the one with $N_f$ flavors, while $N_c$ is fixed.
Notice also that downlifting applies to a generic spectrum of massless fermions without ad-hoc assumptions. In particular, it does not require elements of $\mathcal{R}[N_f-1]$ to be in one-to-one correspondence with elements of $\mathcal{R}[N_f]$, which is something 
the early works in~\cite{tHooft:1979rat, Frishman:1980dq, Farrar:1980sn, Cohen:1981iz, Kaul:1981fd, Takeshita:1981sx} relied upon, see~\cite{Ciambriello:2022wmh} for more explanations.

Finally, notice that the downlifting theorem does not rely on whether QCD$[N_c, N_f]$ confines: it can be considered as a mathematical statement on AMC$[N_f]$ and PMC$[N_f]$, where the latter are defined as a formal system of equations with no a priori relation with the actual dynamics of QCD$[N_c, N_f]$.

\section{Prime factor}
It was observed in~\cite{Preskill:1981sr, Weinberg:1996kr} that for a massless spectrum of baryons there exist no integral solutions of the $[SU(N_f)_{L, R}]^2 U(1)_V$ AMC when $N_c=3$ and $N_f$ is a multiple of $3$, hence $\chi$SB must occur if the confining description applies. 
Motivated by this result, we generalize it by proving the following theorem valid for a generic spectrum:
\\[0.25cm] \textit{In QCD$[N_c, m p]$, where $p$ is a prime factor of $N_c$ and $m$ a positive integer, there exist no integral solutions of the $[SU(m p)_{L, R}]^2 U(1)_V$ AMC. Therefore, $\chi$SB must occur in QCD$[N_c, m p]$ if the theory confines.}
\\[0.3cm]
The proof goes as follows. Let us consider a proper irrep $r=(r_L, r_R, v)$ of $\mathcal{G}[N_f]$, where $r_{L,R}$ are irreps of $SU(N_f)_{L, R}$ and $v$ is the $U(1)_V$ charge. The baryon number $b$ is defined so that $v=b N_c$. The discrete quotient in $\mathcal{G}[N_f]$ leads to the following constraints:  $\mathbb{Z}_{N_c}$ implies that the baryon number $b$ is an integer for color singlets, while $\mathbb{Z}_{N_f}$ implies that $\mathcal{N}(r_L)+ \mathcal{N}(r_R)= v \mod N_f$, where $\mathcal{N}(r_{L,R})$ are the $N_f$-alities of $r_{L,R}$. For $N_f=p$, where $p$ is a prime factor of $N_c$, the above two constraints imply 
\be
\mathcal{N}(r_L)+ \mathcal{N}(r_R)= 0 \mod p\, .
\label{eq:constraint}
\ee
Notice that the hypothesis of confinement is key to obtain this equation, since it ensures that the IR massless states are proper irreps of $\mathcal{G}[N_f]$, and thus have integer baryon number.
There are two possible cases to satisfy Eq.~(\ref{eq:constraint}): either $\mathcal{N}(r_L)=0 \mod p$ (hence $\mathcal{N}(r_R)=0 \mod p$)
or $\mathcal{N}(r_L) \neq 0 \mod p$ (hence $\mathcal{N}(r_R) \neq 0 \mod p$). 
By writing the $[SU(p)_L]^2 U(1)_V$ anomaly coefficient as $A(r)= T(r_L) d(r_R) v$, where $T(r_L)$ is the Dynkin index of $r_L$ and $d(r_R)$ is the dimension of $r_R$, the corresponding AMC$[p]$ reads
\be
1=\sum_{r\in \mathcal{R}[p]} \ell(r)\, T(r_L) d(r_R) b\, .
\ee
This AMC equation cannot be solved for integral values of the $\ell(r)$'s as long as
\be
T(r_L)\, d(r_R)=0 \mod p\ . 
\label{eq:prime_factor}
\ee
In the following, we show that Eq.~(\ref{eq:prime_factor}) holds true by proving that either $T(r_L)=0 \mod p$ or $d(r_R)=0 \mod p$.

First, we show that if $\mathcal{N}(r_L)=0 \mod p$, then $T(r_L)=0 \mod p$. To this aim, we compute the Dynkin index $T(r_L)$ using the generator $T_D=\text{diag}(1,1, ..., -(p-1))$ of $SU(p)_L$, which generates the $U(1)_D$ subgroup.
We notice that the center $\mathbb{Z}_p$, defined as
\be
\mathbb{Z}_p=\{e^{i\frac{2\pi k}{p}T_D}=e^{i\frac{2\pi k}{p}},\;k=0,1,...,p-1\}\ ,
\ee
acts trivially on $r_L$ when $\mathcal{N}(r_L)=0 \mod p$. From another perspective, the $\mathbb{Z}_p$ center is also a subgroup of $U(1)_D$. 
Hence if we decompose $r_L$ into irreps of $U(1)_D$, the corresponding charges $q$ have to satisfy the constraint $e^{i\frac{2\pi}{p}q}=1$, this implies that $q=pn$ with $n$ being an integer. Therefore, when $\mathcal{N}(r_L)=0 \mod p$, the Dynkin index $T(r_L)$ equals
\be
\label{eq:dynkin}
\frac{\text{Tr}_{r_L}[(T_D)^2]}{\text{Tr}_{fund.}[(T_D)^2]} = \frac{\sum_{n} k(r_L \to pn) p^2 n^2}{p(p-1)}= 0 \mod p\ ,
\ee
where the integer $k(r_L \to pn)$ counts how many times the irrep of $U(1)_D$ with charge $pn$ appears in the decomposition of $r_L$. From Eq.~(\ref{eq:dynkin}) it follows that $T(r_L)=0 \mod p$.

Likewise, it is possible to show that if $\mathcal{N}(r_L) \neq 0 \mod p$, then $d(r_R)=0 \mod p$. A proof based on the Weyl dimension formula is given in the Supplementary Material. Therefore, the identity of Eq.~(\ref{eq:prime_factor}) follows.

Finally, we discuss theories with $N_f = mp$ and $m >1$. We consider the subgroup
\begin{equation}
SU(p)_L^m \times SU(p)_R^m \times U(1)_L^{m-1} \times U(1)_R^{m-1} \times U(1)_V
\end{equation}
of ${\cal G}[mp]$ and decompose $(r_L,r_R,v)$ accordingly into irreps
\begin{equation}
\left( r_L^{1},\dots, r_L^{m}, r_R^{1}, \dots , r_R^{m}, q_1, \dots , q_{2 m-2}, v\right) \, ,
\end{equation}
where $r^i$ is an irrep of the $i$-th $SU(p)$ factor, while $q_i$ are the charges under $U(1)_L^{m-1} \times U(1)_R^{m-1}$.
Next, we compute the Dynkin index $T(r_L)$ by using a generator $T_1$ of the first $SU(p)$, and write the anomaly coefficient as
\begin{equation}
  \label{eq:Adec}
  \sum_j T(r_L^{1,j}) d(r_L^{2,j}) \dots d(r_L^{m,j}) d(r_R^{1,j})\dots d(r_R^{m,j}) b\, ,
\end{equation}
where $j$ runs over all irreps in the decomposition. Each term of this sum is a multiple of $p$. Indeed, if ${\cal N}(r_L^{1,j})= 0 \mod p$ then $T(r_L^{1,j}) =0 \mod p$. If instead ${\cal N}(r_L^{1,j})\neq 0 \mod p$, then we notice that Eq.~(\ref{eq:constraint}) implies
\begin{equation}
\sum_{i= 1}^m \left( {\cal N}(r_L^{i,j}) + {\cal N}(r_R^{i,j}) \right) = 0 \mod p\, ,
\end{equation}
which in turn requires that there exists at least one irrep $r_*$ among the remaining $r_L^{i,j}$ and $r_R^{i,j}$ whose $N_f$-ality is non vanishing. Hence $d(r_*) = 0 \mod p$ and the corresponding term in the sum of Eq.~(\ref{eq:Adec}) is proportional to~$p$.
This completes our proof of the theorem.

Combining downlifting and the above result valid for QCD$[N_c, p_{min}]$, we conclude that $\chi$SB must occur in QCD$[N_c, N_f]$ for any number of flavors $N_f \geq p_{min}$ for which the theory confines, where $p_{min}$ is the smallest prime factor of $N_c$.
In QCD ($N_c=3$), in particular, our results imply $\chi$SB for any $N_f\geq 3$ assuming confinement.

\section{Continuity}
It is possible to prove $\chi$SB for $N_f<p_{min}$ if one makes one additional assumption: that of the absence of phase transitions when the quark masses are sent to infinity. The argument is based on continuity and is also valid for a generic spectrum of massless composite fermions.~\footnote{Our line of reasoning is similar to the one used by Vafa and Witten in~\cite{Vafa:1983tf} to prove that isospin is unbroken in the limit of vanishing quark masses; see also~\cite{Preskill:1981sr}.}

Let us consider a theory with $N_f$ massless flavors and $(p_{min}-N_f)$ massive flavors. We denote this theory by $\text{QCD}[N_c,N_f;(p_{min}-N_f)]$. Suppose that the $SU(N_f)_L\times SU(N_f)_R$ chiral symmetry is unbroken by the vacuum for any values of the quark masses in a neighborhood of the origin. This means that the effective potential $V(\phi)$ has a global minimum at $\phi=0$, where $\phi$ is the expectation value of any color-singlet operator which transforms non-trivially under the chiral symmetry. Then, continuity of $V(\phi)$ with respect to the quark masses implies that an $SU(p_{min})_L\times SU(p_{min})_R$ preserving vacuum exists in the limit where all the masses vanish. This is because the vectorlike $SU(p_{min})_V$ symmetry cannot be spontaneously broken~\cite{Vafa:1983tf}, so the unbroken chiral symmetry has to be enhanced to $SU(p_{min})_L\times SU(p_{min})_R$ in order to accommodate both $SU(N_f)_L\times SU(N_f)_R$ and $SU(p_{min})_V$ symmetries.  If QCD$[N_c,p_{min}]$ confines, this contradicts the result obtained previously for $N_f=p_{min}$. Hence, our initial assumption is falsified and $SU(N_f)_L\times SU(N_f)_R$ must be spontaneously broken in $\text{QCD}[N_c,N_f; (p_{min}-N_f)]$.

As a last step, one can send the quark masses to infinity and obtain $\text{QCD}[N_c,N_f]$ from $\text{QCD}[N_c,N_f; (p_{min}-N_f)]$. In the absence of phase transitions, $\chi$SB will persist. This completes our proof.
For QCD with $N_c=3$, it implies that $\chi$SB occurs with $N_f=2$ massless flavors.

\section{Conclusions}
We showed that $\chi$SB must occur in the confining regime of QCD-like theories, thereby elevating the conventional wisdom to the level of a rigorous proof.
Our argument does not rely on dynamical assumptions about the spectrum of massless bound states contrary to previous attempts~\cite{Ciambriello:2022wmh}.
For $N_f \geq p_{min}$, where $p_{min}$ is the smallest prime factor of $N_c$, our proof is algebraic and based on a novel strategy of using AMC and PMC called `downlifting'.
This result only relies on confinement.
For $N_f < p_{min}$ our reasoning makes use of a continuity argument and holds as long as there are no phase transitions when the quark masses are sent to infinity.

These results do not directly apply to theories in the conformal window because in that case a low-energy description in terms of only color-singlet particles does not exist.

\vspace{0.1in}
\noindent \textbf{Acknowledgments.---}
We would like to thank Marco Bochicchio, Ethan Neil, Slava Rychkov, Yael Shadmi and Giovanni Villadoro for useful discussions and comments.
This research was supported in part by the MIUR under contract 2017FMJFMW (PRIN2017), and performed in part at Aspen Center for Physics, which is supported by NSF grant PHY-2210452.
The work of R.C. was partly supported by a grant from the Simons Foundation and by the Munich Institute for Astro- and Particle Physics (MIAPP), which is funded by the DFG (German Research Foundation) under Germany's Excellence Strategy - EXC-2094 – 390783311.
The work of A.L. was partially supported by the INFN special initiative grant GAST.
L.X.X. would like to thank Scuola Normale Superiore in Pisa, where this project was initiated, for its warm hospitality. The work of L.X.X. was partially supported by European Research Council (ERC) grant n.101039756.

\bibliography{bibliografia}

\clearpage
\onecolumngrid
\appendix
\makeatletter

\label{supp}

\newpage

\begin{center}
   \textbf{\large SUPPLEMENTARY MATERIAL \\[.2cm] ``A novel strategy to prove chiral symmetry breaking in QCD-like theories'' }\\[.2cm]
\end{center}

We clarify various technical but useful aspects of our analysis in the following. 
\begin{itemize}
    \item We start by reviewing the global structure of the chiral symmetry group of QCD-like theories, including the discrete identification of the trivial group elements. Similar discussions can be found in~\cite{Tanizaki:2018wtg}.
    \item Next, we present some specific features of the PMC equations, valid for a massless spectrum of baryons, which allow us to downlift with only PMC$[N_f,1]$. As an example, see~\cite{Ciambriello:2024msu} for detailed calculation in QCD$[5, N_f]$ with minimal baryons.  
    \item Then we prove $d(r_R)=0 \mod p$ when $\mathcal{N}(r_L) \neq 0 \mod p$ using the Weyl dimension formula. This completes the proof of $\chi$SB in the confining description of QCD$[N_c,p]$ where $p$ is a prime factor of $N_c$. Some numerical results are also provided. 
\end{itemize}
 
\noindent
\section{Global structure of the QCD symmetry group}
\label{app:discrete_quotient}
\setcounter{equation}{0}
\setcounter{figure}{0}
\setcounter{table}{0}
\renewcommand{\theequation}{A\arabic{equation}}
\renewcommand{\thefigure}{A\arabic{figure}}
\renewcommand{\thetable}{A\arabic{table}}

Let us start by considering the covering group of the internal symmetry group acting on fundamental quarks 
\be
{\tilde{\mathcal{G}}}_{int}[N_f]=SU(N_c) \times SU(N_f)_L \times SU(N_f)_R \times U(1)_V \times \mathbb{Z}_{2N_f} \, ,
\ee
where $SU(N_c)$ is the gauge group and $\mathbb{Z}_{2N_f}$ is the discrete subgroup of $U(1)_A$ which is left unbroken by the ABJ anomaly. Both left-handed and right-handed quarks in QCD-like theories are in the fundamental representation of $SU(N_c)$, which is the reason why QCD-like theories are vectorlike theories. Left-handed quarks $q_L$ are in the fundamental representation of $SU(N_f)_L$ and have charge $+1$ under both $U(1)_V$ and $U(1)_A$; right-handed quarks $q_R$ are in the fundamental representation of $SU(N_f)_R$ and have charge $+1$ under $U(1)_V$ but charge $-1$ under $U(1)_A$. 

One can find the internal symmetry group which acts \emph{faithfully} on fundamental quarks by removing all the trivial group elements in ${\tilde{\mathcal{G}}}_{int}[N_f]$. 
First of all, we notice that $\mathbb{Z}_{2N_f}$ is completely redundant, as one can always undo its action by performing a $U(1)_V$ transformation followed by a transformation of the center of $SU(N_f)_L$:
\be
\begin{array}{c} q_L\;\\ q_R\end{array} 
\xrightarrow{e^{\frac{2\pi i}{2N_f}} \in \mathbb{Z}_{2N_f}}  
\begin{array}{cc} e^{\frac{2\pi i}{2N_f}} &q_L\;\\ e^{-\frac{2\pi i}{2N_f}} & q_R\end{array} 
\xrightarrow{e^{\frac{2\pi i}{2N_f}}\in U(1)_V} 
\begin{array}{cc} e^{\frac{2\pi i}{N_f}} & q_L\;\\ 1 & q_R\end{array} \xrightarrow{e^{-\frac{2\pi i}{N_f}}\in SU(N_f)_L} 
\begin{array}{c} q_L\;\\ q_R\end{array}\;.
\ee 
Furthermore, since any transformation in the centers of $SU(N_c)$ and of the vectorlike $SU(N_f)_V$ can be undone by a $U(1)_V$ transformation, these subgroups also act trivially on quarks. By removing them, we obtain the internal symmetry group acting faithfully on quarks, i.e.
\be
\mathcal{G}[N_f]_{q}=\frac{SU(N_c)\times SU(N_f)_L\times SU(N_f)_R \times U(1)_V}{\mathbb{Z}_{N_c}\times \mathbb{Z}_{N_f}}\;.
\ee
By simply removing the $SU(N_c)$ gauge group, we have the global symmetry group acting faithfully on gauge-invariant color singlets
\bea
\label{eq:GNf}
\mathcal{G}[N_f]=\frac{SU(N_f)_L\times SU(N_f)_R\times U(1)_V}{\mathbb{Z}_{N_c} \times \mathbb{Z}_{N_f}}\ .
\eea

For completeness, we present another derivation of $\mathcal{G}[N_f]$, obtained by operating at the level of color-singlet asymptotic states, which offers a different but equivalent perspective. 
The $SU(N_c)$ gauge group acts trivially on color singlets interpolated by gauge invariant operators, so it needs to be removed first from ${\tilde{\mathcal{G}}}_{int}[N_f]$. However, the rest of ${\tilde{\mathcal{G}}}_{int}[N_f]$ can act nontrivially on color singlets. Let us consider a color singlet interpolated by a gauge-invariant composite operator made of $n_L$ ($n_R$) left (right) quark fields and $\bar{n}_L$ ($\bar{n}_R$) left (right) antiquark fields; the $\mathbb{Z}_{2N_f}$ subgroup of $U(1)_A$ acts on such state as
\be
e^{\frac{2 \pi i}{2N_f} (n_L-n_R-\bar{n}_L+\bar{n}_R)}\, .
\ee
If one combines this trasformation with the following proper rotation of $U(1)_V$
\be
e^{\frac{2 \pi i}{2N_f} (n_L+n_R-\bar{n}_L-\bar{n}_R)}\ ,
\ee
one has the superposition
\be
e^{\frac{2 \pi i}{N_f} (n_L-\bar{n}_L)}\ .
\ee
Clearly, this can be undone by a rotation in the $SU(N_f)_L$ center.
This shows that $\mathbb{Z}_{2N_f}$ in ${\tilde{\mathcal{G}}}_{int}[N_f]$ is also a trivial group when acting on color singlets, and it can be removed. Furthermore, the center of the vectorlike $SU(N_f)_V$ can be removed since its action can be undone by $U(1)_V$, this is the same as at the level of quarks. Finally, we notice that the $U(1)_V$ charge is quantized in integer multiples of $N_c$ when individual quarks have charge $+1$. This is easily seen from the fact that the Young tableaux of $SU(N_c)$ singlets always have a number of boxes which is a multiple of $N_c$. As a result, the $\mathbb{Z}_{N_c}$ subgroup of $U(1)_V$ is also a trivial group acting on color singlets and it needs to be removed. In conclusion, we end with the same $\mathcal{G}[N_f]$ as in Eq.~(\ref{eq:GNf}).

A similar analysis can be performed when some of the quark flavors become massive. When $i$ flavors have non-vanishing and unequal masses, the global symmetry group which acts faithfully on color singlets is
\be
\mathcal{G}[N_f,i]=\frac{SU(N_f-i)_L\times SU(N_f-i)_R \times U(1)_{\hat{V}} \times U(1)_{H_1} \times ... \times U(1)_{H_i}}{\mathbb{Z}_{N_c}\times \mathbb{Z}_{N_f-i}}\ .
\ee
Transformations in the axial $\mathbb{Z}_{2(N_f-i)}$ acting on the massless flavors can be undone by means of $U(1)_{\hat{V}}$ and the center of $SU(N_f-i)_L$, while those in the center $\mathbb{Z}_{N_f-i}$ of the vectorlike $SU(N_f-i)_V$ can be undone by  a proper rotation of $U(1)_{\hat{V}}$; both must be then modded out. Finally, the $\mathbb{Z}_{N_c}$ in the quotient can be viewed as the center of the $SU(N_c)$ gauge group, which can be undone by various rotations of $U(1)$ in $\mathcal{G}[N_f,i]$. Hence it is a trivial group. Alternatively, $\mathbb{Z}_{N_c}$ can be viewed as a trivial subgroup of $U(1)_V$, whose charge equals the sum of the charges under $U(1)_{\hat{V}}, U(1)_{H_1}, \cdots, U(1)_{H_i}$, and it is quantized in multiples of $N_c$ due to gauge invariance, although the smallest charge of each $U(1)$ in $\mathcal{G}[N_f,i]$ is 1.

\noindent
\section{Structure of PMC for baryons}
\label{app:down_baryons}
\setcounter{equation}{0}
\setcounter{figure}{0}
\setcounter{table}{0}
\renewcommand{\theequation}{B\arabic{equation}}
\renewcommand{\thefigure}{B\arabic{figure}}
\renewcommand{\thetable}{B\arabic{table}}

Massless baryons were extensively discussed in early works~\cite{tHooft:1979rat, Frishman:1980dq, Cohen:1981iz, Schwimmer:1981yy, Kaul:1981fd, Takeshita:1981sx}, and can be defined as the states interpolated by composite operators made of only quarks (see for example~\cite{Ciambriello:2022wmh}).
We show that, for purely baryonic spectra, $\text{PMC}[N_f-1,1]$ is a subset of $\text{PMC}[N_f,1]$, namely any solution of $\text{PMC}[N_f,1]$ automatically solves $\text{PMC}[N_f-1,1]$.
By induction, $\text{PMC}[N_f-2,1]$ is also a subset of $\text{PMC}[N_f-1,1]$ for baryons, etc. Due to the identification between $\text{PMC}[N_f-1,i-1]$ and $\text{PMC}[N_f,i]$ of Eq.~(\ref{eq:identification_PMC}), this implies that all $\text{PMC}[N_f,i]$ with $2\leq i \leq N_f-2$ are subsets of $\text{PMC}[N_f,1]$ for baryons. As a consequence, it is possible to downlift baryonic spectra using only $\text{PMC}[N_f,1]$. 

Let us consider a baryon interpolated by a composite operator made of $n_L$ left-handed and $n_R$ right-handed quark fields and with $U(1)_V$ charge $v$, where 
\be
n_L+n_R=v\, . 
\ee
An irrep characterizing such state  can be denoted as 
\be
r=(\{n_L\},\{n_R\},v)\, ,
\ee
where $\{n\}$ means a Young tableau (YT) with $n$ boxes (here we follow the notation in~\cite{Ciambriello:2022wmh}).
If $v\geq N_f$, then $\{n_L\}$ and $\{n_R\}$ can have columns with $N_f$ boxes, which transform as singlets under $SU(N_f)_{L/R}$. The same irrep $r$ can thus be interpolated by composite operators that differ by groups of $N_f$ fully-antisymmetrized indices (singlets);  any two such operators therefore transform as equivalent tensors~\cite{Ciambriello:2022wmh}.

When one flavor is given a finite mass, any irrep $r$ can be decomposed into a direct sum of irreps $r'$ of $\mathcal{G}[N_f,1]$; these latter can be denoted as
\be
r^\prime=(\{n_L^\prime\},\{n_R^\prime\}, \hat{v}, H_1)\, ,
\ee
where $\hat v$ and $H_1$ are the charges under respectively $U(1)_{\hat V}$ and $U(1)_{H_1}$, and
\be
n_L^\prime+n_R^\prime=\hat{v}\ , \quad\quad \hat{v}+H_1=v\ . 
\ee

There exists one PMC$[N_f,1]$ equation for each $r^\prime$ with non-zero $H_1$ charge. The same irrep $r'$ can correspond to different though equivalent tensors of $\mathcal{G}[N_f,1]$ as long as $\hat v \geq N_f-1$. Since the minimal non-zero value of $H_1$ is~1 and $\hat v = v - H_1$, it is easy to see that the condition $v < N_f$ ensures the absence of equivalent tensors also after decomposition.
Notice that if either $\{n_L\}$ or $\{n_R\}$ in $r$ has $N_f$ rows, then $r$ gets decomposed into irreps $r^\prime$ with $H_1>0$. On the other hand,  irreps $r^\prime$ with $H_1=0$ necessarily have the same YTs as the ones of their parent irrep $r$, i.e. $\{n_L^\prime\}=\{n_L\}$ and $\{n_R^\prime\}=\{n_R\}$.

With these considerations in mind, it is useful to classify PMC$[N_f,1]$ into two types (see Fig.~\ref{fig:abPMC}):
\begin{figure}[t]
\centering
\includegraphics[scale=0.35]{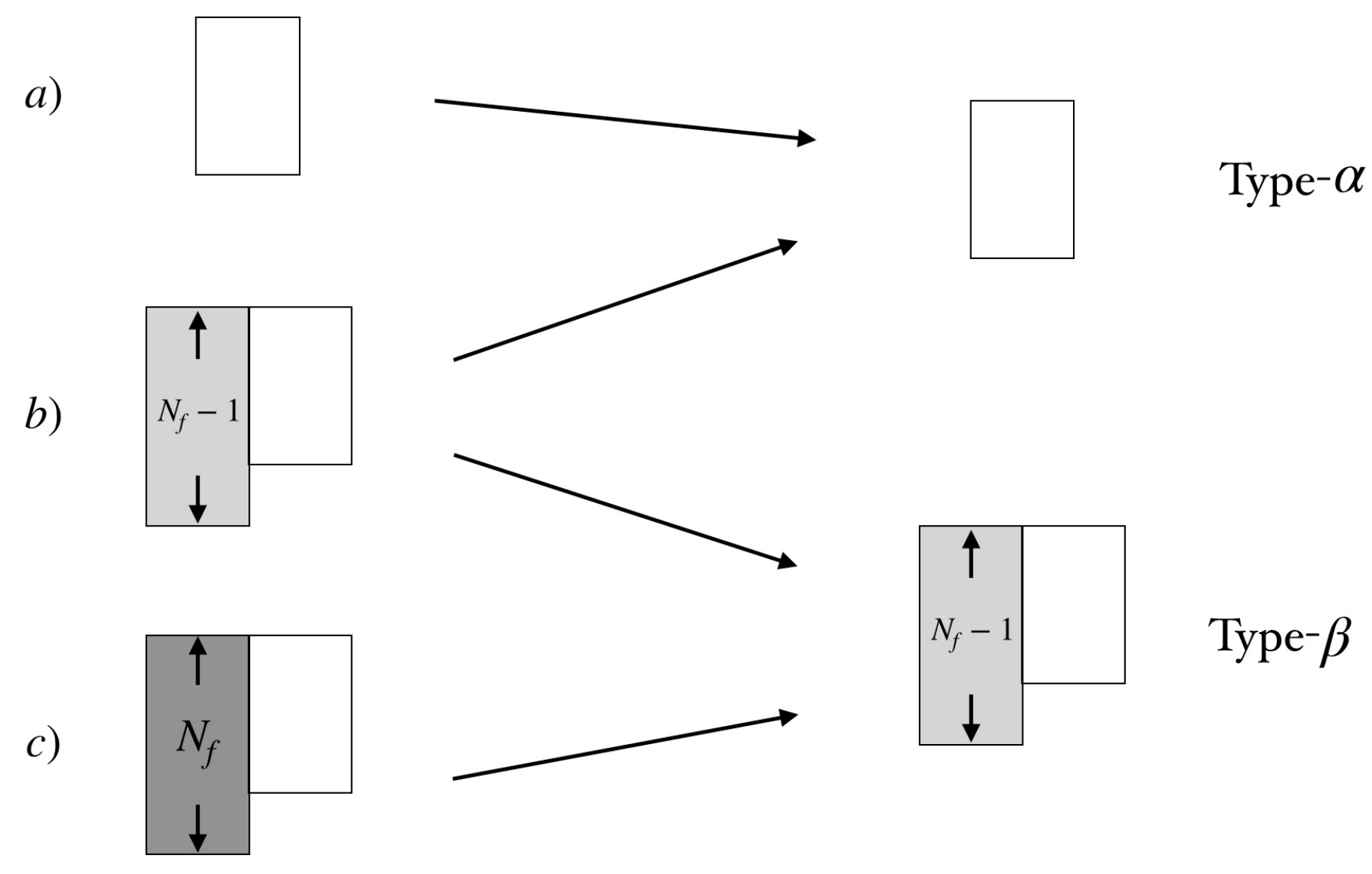}
\caption{Possible contributions to type-$\alpha$ and type-$\beta$ PMC$[N_f,1]$ equations. The Young Tableaux $a)$, $b)$ and $c)$ schematically denote $\{n_L\}$ and $\{n_R\}$ of $r$ with less than $N_f-1$ rows, with $N_f-1$ rows, and with $N_f$ rows, respectively.}
\label{fig:abPMC}
\end{figure}
\begin{enumerate}
	\item Type-$\alpha$ equations are those which equate to zero the indices of $r^\prime$ where both $\{n^\prime_L\}$ and $\{n^\prime_R\}$ have less than $N_f-1$ rows. The baryonic states in $r$ contributing to these equations are those whose YTs $\{n_L\}$ and $\{n_R\}$ have $N_f-1$ rows or less.
	\item Type-$\beta$ equations equate to zero the indices of $r^\prime$ where at least one of the YTs $\{n^\prime_L\}$, $\{n^\prime_R\}$ has $N_f-1$ rows. Only irreps $r$ where at least one of the YTs $\{n_L\}$, $\{n_R\}$ has exactly $N_f$ or $N_f-1$ rows can contribute to this type of equations.
\end{enumerate}
Likewise, one can classify $\text{PMC}[N_f-1,1]$ by replacing $N_f$ with $N_f-1$ everywhere in the definitions. 

By comparison, we find that $\text{PMC}[N_f-1,1]$ is a subset of $\text{PMC}[N_f,1]$. Indeed:
\begin{enumerate}
\item If both YTs in $r'$ have less than $N_f-2$ rows, then for a given type-$\alpha$ $\text{PMC}[N_f,1]$ equation there exists a corresponding type-$\alpha$ $\text{PMC}[N_f-1,1]$ equation. The correspondence is one-to-one and the two equations are identical.
\item If at least one of the YTs in $r'$ has exactly $N_f-2$ rows, then for a given type-$\alpha$ $\text{PMC}[N_f,1]$ equation there exists a corresponding type-$\beta$ $\text{PMC}[N_f-1,1]$ equation. In this case, two or more equations of $\text{PMC}[N_f,1]$ can collapse to the same equation of $\text{PMC}[N_f-1,1]$, i.e. the latter is given by the sum of the former equations. This happens when the irreps $r'$ of the $\text{PMC}[N_f,1]$ equations differ by the position of columns with $N_f-2$ boxes.
 \item Type-$\beta$ equations of $\text{PMC}[N_f,1]$ have no counterpart in $\text{PMC}[N_f-1,1]$. 
\end{enumerate}
The collapse of two or more equations of $\text{PMC}[N_f,1]$ into the same equation of $\text{PMC}[N_f-1,1]$ happens because inequivalent tensors of $SU(N_f-1)_L\times SU(N_f-1)_{R}$ become equivalent tensors of $SU(N_f-2)_L\times SU(N_f-2)_{R}$. There is another consequence of the presence of equivalent tensors. As we already said, in the case of baryons, two tensors of $SU(N_f)_L\times SU(N_f)_{R}$ can be equivalent only because their YTs have $N_f$ rows. When decreasing the number of flavors, these tensors are not well defined anymore. This means that some of the representations in $\mathcal{R}[N_f]$ will not exist in $\mathcal{R}[N_f-1]$. Notice however that irreps $r$ whose YTs have $N_f$ rows contribute only to type-$\beta$ $\text{PMC}[N_f,1]$ equations, and these equations do not have a counterpart in $\text{PMC}[N_f-1,1]$. In conclusions: some of the variables (indices) disappear when decreasing the number of flavors, while the remaining ones are subject to a subset of the original equations.

An explicit example which illustrates the structure of baryonic PMC is given in~\cite{Ciambriello:2024msu}.

\noindent
\section{Proof of $\chi$SB in QCD$[N_c,p]$ for $\mathcal{N}(r_R) \neq 0 \mod p$}
\label{app:prime_factor}
\setcounter{equation}{0}
\setcounter{figure}{0}
\setcounter{table}{0}
\renewcommand{\theequation}{C\arabic{equation}}
\renewcommand{\thefigure}{C\arabic{figure}}
\renewcommand{\thetable}{C\arabic{table}}
In this Appendix we show that if $\mathcal{N}(r_R) \neq 0 \mod p$ (hence $\mathcal{N}(r_L) \neq 0 \mod p$) then $d(r_R)=0 \mod p$. This completes our proof of $\chi$SB in QCD$[N_c,p]$, where $p$ is a prime factor of $N_c$, discussed in the main text.

Let us start by writing the Weyl dimension formula for an irrep $r_R$ of $SU(N_f)_R$ with highest weight $\Lambda=(\Lambda_1,...,\Lambda_{N_f-1})$ (see for example~\cite{Yamatsu:2015npn} and references therein):
\be
\label{eq:drR}
d(r_R)=\prod_{i=2}^{N_f}\prod_{h=1}^{i-1}\left(\frac{\sum_{l=h}^{i-1}\Lambda_{l}}{i-h}+1\right)\ .
\ee
The Young tableau corresponding to $r_R$ features $\Lambda_i$ columns with $i$ boxes, hence the total number of boxes equals $\mathcal{N}(r_R)=\sum_{i=1}^{N_f-1}i\cdot \Lambda_i$. (Clearly, columns with $N_f$ boxes are not included in this definition.) Alternatively, the same Young tableau can be defined using a partition $a=(a_1,...,a_k)$ of the $\mathcal{N}(r_R)$ boxes, where $a_i$ is the number of boxes in the $i$-th row of the Young tableau, such that $\mathcal{N}(r_R)=\sum_{i=1}^{k} a_i$. Here, $k$ is the largest integer for which $a_k\not = 0$, with $k\leq N_f-1$, i.e. the number of rows of the Young tableau (height of the diagram).
The relation between the above two definitions is that $\Lambda_i=a_i-a_{i+1}$ for $i<k$, $\Lambda_k=a_k$ and $\Lambda_i=0$ for $i>k$. 

To rewrite $d(r_R)$ using the partition $a$, we need to consider the following three cases.
\begin{enumerate}
    \item When $2\leq i\leq k$ (and $h<i$), we have $\sum_{l=h}^{i-1}\Lambda_{l}=a_h-a_i$. The numerator of $d(r_R)$ receives the contribution 
    \be
	N_1=\prod_{i=2}^{k}\prod_{h=1}^{i-1}\left[a_h-a_i+i-h\right]\ ,
	\ee
    which can be rewritten as
    \be
    N_1=\prod_{H=1}^{k-1}\prod_{I=H+1}^{k}\left[a_{k+1-I}-a_{k+1-H}+I-H\right]
    \ee
    by changing variables $i=k+1-H$ and $h=k+1-I$. If we define $q_I=a_{k+1-I}+I$, then
    \be
    N_1=\prod_{H=1}^{k-1}\prod_{I=H+1}^{k}(q_{I}-q_{H})=\prod_{I=2}^{k}\prod_{H=1}^{I-1}(q_{I}-q_{H})\ .
    \ee
    At the same time, the denominator of $d(r_R)$ receives the contribution
    \be
	D_1=\prod_{i=2}^{k}\prod_{h=1}^{i-1}(i-h)=\prod_{i=2}^{k}(i-1)!=\prod_{i=1}^{k-1}i!\ .
    \ee
    Notice that the above terms are present only when $k>1$.
    \item When $h\leq k$ and $k+1\leq i\leq N_f$, we have $\sum_{l=h}^{i-1}\Lambda_{l}=a_h$. The numerator of $d(r_R)$ receives the contribution
    \be
	N_2=\prod_{i=k+1}^{N_f}\prod_{h=1}^{k}\left[a_h+i-h\right]
	\ee
	If we change variables $N_f=k+1+j$, $i=k+1+H$ and $h=k+1-I$, we get
	\be
	N_2=\prod_{H=0}^{j}\prod_{I=1}^{k}\left[a_{k-I+1}+I+H\right]=\prod_{I=1}^{k}\prod_{H=0}^{j}\left[q_I+H\right]\ ,
	\ee
    where $q_I=a_{k+1-I}+I$. At the same time, the denominator of $d(r_R)$ receives the contribution
    \be
	D_2=\prod_{i=k+1}^{k+j+1}\prod_{h=1}^{k}(i-h)=\prod_{i=k+1}^{k+j+1}(i-1)...(i-k)=\prod_{i=k}^{k+j}i...(i-k+1)=\prod_{i=k}^{k+j}\frac{i!}{(i-k)!}\, .
	\ee
      \item When $k<h<i$ and $k+1\leq i\leq N_f$, we have $\sum_{l=h}^{i-1}\Lambda_{l}=0$, hence the quantity in parenthesis in Eq.~(\ref{eq:drR}) is equal to 1. We can therefore just neglect this case.
\end{enumerate}

To summarize, we find the following expression for the full denominator $D$:
\be
\begin{split}
  \label{eq:D}
D=D_1 D_2&=\left[\prod_{i=k}^{k+j}\frac{i!}{(i-k)!}\right]\cdot
\begin{cases}
\prod_{i=1}^{k-1}i!\;\;&\text{when }k>1\\
1&\text{when }k=1
\end{cases}
=\frac{\prod_{i=1}^{k+j}i!}{\prod_{i=k}^{k+j}(i-k)!}=\frac{\prod_{i=1}^{k+j}i!}{\prod_{i=0}^{j}i!}\\
&=\prod_{i=j+1}^{k+j}i!=\prod_{i=1}^{k}(i+j)!\ \ .
\end{split}
\ee
Therefore, we have $d(r_R)=(N_2/D)\cdot N_1$ when $k>1$ and $d(r_R)=N_2/D$ when $k=1$, i.e.
\be
d(r_R)=\left[\prod_{i=1}^{k}\frac{\prod_{h=0}^{j}(q_i+h)}{(i+j)!}\right]\cdot
\begin{cases}
\prod_{i=2}^{k}\prod_{h=1}^{i-1}(q_i-q_h)\;\;&\text{when }k>1\\
1&\text{when }k=1
\end{cases}\ ,
\label{eq:weyl}
\ee
where $N_f=k+1+j$ and $q_i=a_{k+1-i}+i$, with $i$ ranging from $1$ to $k$.

Let us now take $N_f=p$, where $p$ is a prime factor of $N_c$, and consider the case in which $\mathcal{N}(r_R) \neq 0 \mod p$. We want to prove that $d(r_R)=0 \mod p$. Clearly, the prime factor $p$ cannot appear in the full denominator~$D$, see Eq.~(\ref{eq:D}). All we have to show is that either $N_1=0 \mod p$ or $N_2=0 \mod p$. 

When $k=1$, we have $q_1=a_1+1=\mathcal{N}(r_R)+1$, hence
\be
N_2=(\mathcal{N}(r_R)+1) (\mathcal{N}(r_R)+2) \cdots (\mathcal{N}(r_R)+p-1). 
\ee
Suppose that $\mathcal{N}(r_R) = s \mod p$ and $0<s<p$, then $\mathcal{N}(r_R) +(p-s) = 0 \mod p$. Since the factor $\mathcal{N}(r_R) +(p-s)$ is also contained in $N_2$, it follows that $N_2=0 \mod p$. This completes the proof for $k=1$. 

When $k>1$, we can start by assuming $N_1\neq 0 \mod p$ and $N_2\neq 0 \mod p$, and show by contradiction that these two conditions cannot simultaneously be satisfied.  Let us define $s_i=q_i \mod p$. 
When $N_1\neq 0 \mod p$, it implies that $s_1, s_2,\cdots,s_k$ are all different from each other. If any $s_i$ among $s_1, s_2,\cdots,s_k$ vanishes, then $N_2= 0 \mod p$; likewise, if any $s_i\geq k+1$, then $0<p-s_i\leq j$ and in the product of $N_2$ there is a factor $q_i+(p-s_i)=0 \mod p$. Again, it follows that $N_2= 0 \mod p$. Hence, to satisfy both $N_1\neq 0 \mod p$ and $N_2\neq 0 \mod p$, the set $\{s_1, s_2,\cdots,s_k\}$ is necessarily a permutation of $\{1,2,\cdots,k\}$, and it follows that
\be
\sum_{i=1}^{k} s_i=\frac{(k+1)k}{2} \mod p. 
\ee
From another perspective, 
\be
\sum_{i=1}^{k} s_i=\sum_{i=1}^{k} q_i \mod p=\left(\sum_{i=1}^{k} a_i +\sum_{i=1}^{k} i\right)  \mod p=\mathcal{N}(r_R)+\frac{(k+1)k}{2} \mod p\ . 
\ee
The above two results contradict each other since $\mathcal{N}(r_R) \neq 0 \mod p$. This completes the proof for $k>1$. 

We end this Appendix by providing some numerical examples to support the results that we derived for QCD$[N_c,p]$:
\begin{itemize}
\item In Table~\ref{numeric1}, we list the irreps of $SU(N_f=3)$ with up to $n=7$ boxes in their Young tableau, and show their Dynkin indices and dimensions. It can be seen that $T(r)=0 \mod p$ when $n=0 \mod p$, and $d(r)=0 \mod p$ when $n\neq 0 \mod p$.
\item Table~\ref{numeric2} reports the values of $N_1$, $N_2$, and $s_i$ for the  irreps of $SU(N_f=5)$ with $5$ and $6$ boxes in their Young tableau. All the features encountered in the proof of this Appendix are explicitly verified.
\end{itemize}

\begin{table}[h]
	\small
	\centering
	\ytableausetup{smalltableaux}
	\ytableausetup{aligntableaux=bottom}
	\renewcommand{\arraystretch}{1.7}
	\begin{tabular}[t]{|c|c|cc|}
		\hline
		$n$ &  YT &  Dynkin &  Dimension\\
		\hline\rule{0pt}{3ex}\multirow{1}{*}{1}&$\ydiagram{1}$&$1$&$\textcolor{red}{3}$\\
		\hline\rule{0pt}{5ex}\multirow{2}{*}{2}
        &$\ydiagram{1, 1}$&$1$&$\textcolor{red}{3}$\\
		&$\ydiagram{2}$&$5$&$\textcolor{red}{6}$\\
		\hline\rule{0pt}{7ex}\multirow{3}{*}{3}
        &$\ydiagram{1, 1, 1}$&$\textcolor{red}{0}$&$1$\\
		&$\ydiagram{2, 1}$&$\textcolor{red}{6}$&$8$\\
		&$\ydiagram{3}$&$\textcolor{red}{15}$&$10$\\
		\hline\rule{0pt}{7ex}\multirow{4}{*}{4}
        &$\ydiagram{2, 1, 1}$&$1$&$\textcolor{red}{3}$\\
		&$\ydiagram{2, 2}$&$5$&$\textcolor{red}{6}$\\
		&$\ydiagram{3, 1}$&$20$&$\textcolor{red}{15}$\\
		&$\ydiagram{4}$&$35$&$\textcolor{red}{15}$\\
		\hline\rule{0pt}{7ex}\multirow{5}{*}{5}
        &$\ydiagram{2, 2, 1}$&$1$&$\textcolor{red}{3}$\\
		&$\ydiagram{3, 1, 1}$&$5$&$\textcolor{red}{6}$\\
		&$\ydiagram{3, 2}$&$20$&$\textcolor{red}{15}$\\
		&$\ydiagram{4, 1}$&$50$&$\textcolor{red}{24}$\\
		&$\ydiagram{5}$&$70$&$\textcolor{red}{21}$\\
		\hline
	\end{tabular}
	\quad
	\begin{tabular}[t]{|c|c|cc|}
		\hline
		$n$ &  YT &  Dynkin &  Dimension\\
		\hline\rule{0pt}{7ex}\multirow{7}{*}{6}&$\ydiagram{2, 2, 2}$&$\textcolor{red}{0}$&$1$\\
		&$\ydiagram{3, 2, 1}$&$\textcolor{red}{6}$&$8$\\
		&$\ydiagram{4, 1, 1}$&$\textcolor{red}{15}$&$10$\\
		&$\ydiagram{3, 3}$&$\textcolor{red}{15}$&$10$\\
		&$\ydiagram{4, 2}$&$\textcolor{red}{54}$&$\textcolor{red}{27}$\\
		&$\ydiagram{5, 1}$&$\textcolor{red}{105}$&$35$\\
		&$\ydiagram{6}$&$\textcolor{red}{126}$&$28$\\
		\hline\rule{0pt}{7ex}\multirow{8}{*}{7}&$\ydiagram{3, 2, 2}$&$1$&$\textcolor{red}{3}$\\
		&$\ydiagram{3, 3, 1}$&$5$&$\textcolor{red}{6}$\\
		&$\ydiagram{4, 2, 1}$&$20$&$\textcolor{red}{15}$\\
		&$\ydiagram{5, 1, 1}$&$35$&$\textcolor{red}{15}$\\
		&$\ydiagram{4, 3}$&$50$&$\textcolor{red}{24}$\\
		&$\ydiagram{5, 2}$&$119$&$\textcolor{red}{42}$\\
		&$\ydiagram{6, 1}$&$196$&$\textcolor{red}{48}$\\
		&$\ydiagram{7}$&$\textcolor{red}{210}$&$\textcolor{red}{36}$\\
		\hline
	\end{tabular}
	\caption{Dynkin indices and dimensions of the irreps of $SU(N_f=3)$ with up to 7 boxes in their Young tableau. The numbers in red are multiples of $N_f=3$.}
	\label{numeric1}
\end{table}

\begin{table}[h]
	\ytableausetup{smalltableaux}
	\ytableausetup{aligntableaux=bottom}
	\centering
	\renewcommand{\arraystretch}{1.7}
	\begin{tabular}[t]{|c|c|c|c|}
		\hline
		YT &  $N_1$ &  $N_2$ & $s_i$\\
		\hline\rule{0pt}{11ex}
		$\ydiagram{2, 1, 1, 1}$&$48$&$144$&\textcolor{blue}{$1,\; 2,\;3,\;4$}\\
		$\ydiagram{2, 2, 1}$&$6 $&$\textcolor{red}{3600}$&$0, \; 2,\;4$\\
		$\ydiagram{3, 1, 1}$&$12$&$3024$&\textcolor{blue}{$1,\;2,\;3$}\\
		$\ydiagram{3, 2}$&$2$&$\textcolor{red}{12600}$&$0,\;3$\\
		$\ydiagram{4, 1}$&$4$&$8064$&\textcolor{blue}{$1,\;2$}\\
		$\ydiagram{5}$&$1$&$3024$&\textcolor{blue}{$1$}\\
		\hline
	\end{tabular}
	\quad
	\begin{tabular}[t]{|c|c|c|c|}
		\hline
		YT &  $N_1$ &  $N_2$ & $s_i$\\
		\hline\rule{0pt}{11ex}
		$\ydiagram{2, 1, 1, 1, 1}$&$1$&$\textcolor{red}{120}$&$2$\\
		$\ydiagram{2, 2, 1, 1}$&$72 $&$\textcolor{red}{180}$&$0,\;1,\;2,\;3$\\
		$\ydiagram{3, 1, 1, 1}$&$\textcolor{red}{120} $&$168$&$2,\;2,\;3,\;4$\\
		$\ydiagram{2, 2, 2}$&$2 $&$\textcolor{red}{7200}$&$0,\;3,\;4$\\
		$\ydiagram{3, 2, 1}$&$16 $&$\textcolor{red}{5040}$&$1,\;2,\;4$\\
        $\ydiagram{4, 1, 1}$&$\textcolor{red}{20}$&$4032$&$2,\;2,\;3$\\
		$\ydiagram{3, 3}$&$1$&$\textcolor{red}{25200}$&$0,\;4$\\
		$\ydiagram{4, 2}$&$3$&$\textcolor{red}{20160}$&$1,\;3$\\
		$\ydiagram{5, 1}$&$\textcolor{red}{5}$&$12096$&$2,\;2$\\
		$\ydiagram{6}$&$1$&$\textcolor{red}{5040}$&$2$\\
		
		\hline
	\end{tabular}
	\caption{Irreps of $SU(N_f=5)$ with five and six boxes in their Young tableau, together with the corresponding values of $N_1$, $N_2$, $s_i$. The numbers in red are multiples of $N_f=5$; the $s_i$ in blue are permutations of $1,...,k$.}
	\label{numeric2}
\end{table}

\end{document}